\documentclass{article}
\usepackage[dvips]{graphics}
\usepackage{wrapfig}
\usepackage{epsfig}

\usepackage{romp33,amsmath}

 1   
\font\ddpp=msbm8  scaled \magstep 1  

\def\QED{\hskip0.1em\hfill\null\ \null\nobreak\hfill
\kern3pt\lower1.8pt\vbox{\hrule\hbox   {\vrule\kern1pt\vbox{\kern1.7pt
\hbox{$\scriptstyle   QED$}\kern0.2pt}\kern1pt\vrule}\hrule}}
\def\R{\hbox{\ddpp R}}               

\newcounter{proMM}
\newcounter{remMM}
\newenvironment{proposition}[1]{\refstepcounter{proMM}\trivlist
   \item[\hskip19pt{\sc #1~\arabic{proMM}.}]\it\hskip3pt}{\endtrivlist}
\newenvironment{remark}[1]{\refstepcounter{remMM}\trivlist
   \item[\hskip19pt{\sc #1~\arabic{remMM}.}]\it\hskip3pt}{\endtrivlist}

\title{Nonholonomic constraints in classical field theories}
\author{ Ernst Binz\\
Fakult\"at f\"ur Mathematik und Informatik\\
Universit\"at Mannheim,
68131 Mannheim, Germany\\
binz@math.uni-mannheim.de\\[2ex]
Manuel de Le\'on and David Mart{\'\i}n de Diego
\thanks{Partially supported by grants DGICYT (Spain) PB97-1257
and PGC2000-2191-E, and the University of Mannheim.}\\ 
IMAFF,
Consejo Superior de Investigaciones Cient{\'\i}ficas\\
Serrano 123, 28006 Madrid, Spain\\ mdeleon@imaff.cfmac.csic.es
and d.martin@imaff.cfmac.csic.es \\[2ex]
Dan Socolescu\\
Fachbereich Mathematik, 
Universit\"at Kaiserslautern\\
67663 Kaiser\-lau\-tern, Germany\\
socolescu@mathematik.uni-kl.de }

\begin{document}

\maketitle
\begin{abstract}
     A multisymplectic setting for classical field theories
subjected to non-holonomic constraints is presented. The infinite dimensional 
setting in the space of Cauchy data is also given.
 \end{abstract}


\section{Introduction}

There is a renewal interest in two, in principle, different
topics. One is the multisymplectic formalism for classical field theories,
introduced in the latest seventies by the Polish school led by Tulczyjew 
(see \cite{BSF} for details),
and independently, by P.L. Garc{\'\i}a {\it et al} \cite{PLG1,PLG2}, and
H. Goldsmichdt and Sh. Sternberg \cite{GoSt} 
(see \cite{CIL1,CIL2,bar2,bar3,gimmsy1,gimmsy2,hrabak1,hrabak2,paufler} for
recent approaches).
The other topic is mechanical systems subjected
to non-holonomic constraints. This last topic, being a classical subject in
the literature, has recently become very prominent in the area of
Geometric Mechanics.

Our aim in this paper is to give a multisymplectic setting for
first-order field theories subjected to non-holonomic constraints, that is,
the lagrangian depends on the first derivatives of the fields, and, in addition,
not all the values are allowable.
We do not know interesting examples in physics:
for instance, $\hbox{div} \; {\bf E}=0$ in electromagnetism, 
or $\hbox{div} \; {\bf E}=\rho$ in electromagnetism interacting with 
charged matter (see \cite{MPSW}) are non-holonomic constraints,
but they are dinamically invariant, so that
nothing new can be added using our formalism.
Continuum Mechanics seems to be the main and most interesting source of examples.
 
In \cite{MPSW} the authors consider incompressibility constraint
in fluids and elastic solids, but this type of constraints become holonomic
after integration in the space of Cauchy data.
A genuine type of constraints are exhibited by an elastic body rolling
with constant velocity,
while deforming (for instance, an automobile tire on
pavement \cite{NF}). Another important family of constraints are those appearing
in the dynamics of media with microstructure.
Let us recall that a simple body is mathematically modelled as a three-dimensional
manifold $B$ which can be covered by just one chart; therefore, a configuration is just
an embedding of $B$ into $\R^3$, and a deformation is a change of configuration.
A medium with microstructure is more involved: it is modelled as a principal
bundle $P \longrightarrow B$ which is embedded in a physical ambient
principal bundle. The configurations and the deformations are 
now principal bundle isomorphisms \cite{BLS1}.
A multisymplectic setting for elastic media
was developed in \cite{MdeL}, and it is a natural 
continuation of the nice work in \cite{MH}.
Many of these continuum systems exhibit kinematic constraints
given by the existence of a connection in the ambient 
principal bundle \cite{BLS2,Capriz}.

In this paper, apart from presenting a multisymplectic setting for first-order field theories
subjected to non-holonomic constraints, we also develop the corresponding
infinite dimensional setting in the space of Cauchy data, proving that the constraint
forces can be integrated over the Cauchy surface to get
a constrained 2-form on the space of Cauchy data. So, the
equations of motion are obtained modifying the usual 2-form with this
constrained 2-form.

\section{Multisymplectic geometry}

In this section we will recall some fundamentals
on multisymplectic manifolds which will be needed later.
We refer to \cite{CIL1} for more details.

\begin{definition}{Definition}
A closed $k$-form $\omega$ on a manifold $M$ is called {\sl multisymplectic} if
the mapping
$$
X \in T_xM \mapsto i_X \, \omega \in (\Lambda^{k-1}E)_x
$$
is injective for all $x \in M$. In this case, the pair $(M, \omega)$
is called a {\sl multisymplectic manifold}.
\end{definition}

As is well-known, cotangent bundles are the canonical models for symplectic
manifolds. For multisymplectic manifolds we have more variety of models (see \cite{CIL1})
but we are mainly interested in the following ones.

Let $\Lambda^kE$ be the total space of the bundle of $k$-forms
on a manifold $E$. We define a $k$-form on $\Lambda^kE$ by setting
$$
\Theta (\alpha) (X_1, \dots, X_k) = \alpha (T\rho^k_E(\alpha)X_1, \dots,
T\rho^k_E(\alpha) X_k ),
$$
for $\alpha \in (\Lambda^k E)_x$ and $X_1, \dots, X_k \in T_\alpha(\Lambda^kE)$,
where $\rho^k_E : \Lambda^k E \longrightarrow E$ is the canonical projection.

The form $\Omega = - d\Theta$ is a multisymplectic $(k+1)$-form on $\Lambda^kE$.
Of course, for $k=1$ we obtain the canonical symplectic form $\omega_E$ on
$\Lambda^1E=T^*E$.

More interesting models can be defined if $E$ is fibered over a manifold $N$,
say $\pi : E \longrightarrow N$ is a fibration. Indeed, we consider
the bundle of $r$-semibasic forms on $E$:
$$
\Lambda^k_r E = \{\alpha \in \Lambda^kE \; | \;
i_{v_1 \wedge \cdots \wedge v_r} \, \alpha = 0, \; \hbox{for} \;
r \; \pi-\hbox{vertical 
tangent vectors} \; v_1, \dots , v_r \}.
$$

We can reproduce the above definition to obtain a $k$-form $\Theta_r$ on
$\Lambda^k_r E$ which is nothing but the restriction of $\Theta$
to the submanifold $\Lambda^k_rE$. Therefore, the $(k+1)$-form
$\Omega_r = - d\Theta_r$ is again multisymplectic and, of course,
the restriction of $\Omega$.

\section{Multisymplectic formalism for classical field theories}

\subsection{The finite dimensional setting}

The mathematical ingredients of a classical field theory are the following ones:

\begin{itemize}

\item a fibration $\pi_{XY} : Y \longrightarrow X$, where $X$
represents a $n$-dimensional space-time, that is, an oriented
manifold of dimension $n$ with volume form $\eta$;

\item a lagrangian form $\Lambda : Z \longrightarrow \Lambda^n X$,
where $Z$ denotes the 1-jet prolongation $J^1 \pi_{XY}$.

\end{itemize}

There are two induced fibrations $\pi_{XZ} : Z \longrightarrow X$ and
$\pi_{XY} : Y \longrightarrow X$ such that $\pi_{XZ} = \pi_{XY} \circ \pi_{YZ}$.

We have $\Lambda = L \eta$, with
the obvious identifications, and $L : Z \longrightarrow \R$ is the {\sl lagrangian
function}. 

In this paper we will choose fibered coordinates 
$(x^\mu, y^i, z^i_\mu)$ on $Z$ such that
$\eta = d^nx = dx^1 \wedge \dots \wedge dx^n$. A useful notation will be
$d^{n-1}x^i= i_{\frac{\partial}{\partial x^i}}\eta$.

Therefore, the lagrangian function is $L = L(x^\mu, y^i, z^i_\mu)$.
$L$ is said to be {\sl regular} if the hessian matrix
$$
\left(\frac{\partial^2 L}{\partial z^i_\mu \partial z^j_\nu}\right)
$$
is regular. 

We define the {\sl action integral} 
$$
A(\sigma) = \int_U \, (j^1\sigma)^* \Lambda
$$
where $\sigma$ is a section of $\pi_{XY}$ defined on an open set $U$, and
$j^1 \sigma$ denotes its first jet prolongation. A section $\sigma$ is called an
{\sl extremal} for the above action if
$$
\frac{d}{dt} A(\phi_t \circ \sigma)_{|t=0} = 0
$$
for every flow $\phi_t$ on $Y$ such that $\pi_{XY} \circ \phi_t = \pi_{XY}$,
and $\phi_t(y)=y$ for all $y$ in the boundary of $\sigma(U)$.
Since such a flow $\phi_t$ is generated by a $\pi_{XY}$-vertical
vector field $\xi_Y$ vanishing on the boundary of $\sigma(U)$, we then
conclude that $\sigma$ is an extremal if and only if
\begin{equation}\label{cero}
\int_U \, (j^1\sigma)^* {\cal L}_{\xi_Z} \Lambda = 0 ,
\end{equation}
for all $\xi_Y$ satisfying the above conditions, where $\xi_Z$
is the natural prolongation of $\xi_Y$ to $Z$.
Let us recall that if
$$
\xi_Y = \xi^i (x,y) \, \frac{\partial}{\partial y^i},
$$
then
$$
\xi_Z = \xi^i \, \frac{\partial}{\partial y^i} + \frac{\partial \xi^i}{\partial x^\mu} \,
\frac{\partial}{\partial z^i_\mu}.
$$
Therefore, we deduce that $\sigma(x^\mu) = (x^\mu, \sigma^i(x))$ is an 
extremal if and only if
$$
\int_U \, \left[\frac{\partial L}{\partial y^i} - 
\frac{\partial}{\partial x^\mu}(\frac{\partial L}{\partial z^i_\mu})\right]
\, \xi^i \, dx^n = 0
$$
for all values of $\xi^i$. Thus, $\sigma$ will be an extremal if and only if
\begin{equation}\label{EL}
\frac{\partial L}{\partial y^i} - 
\frac{\partial}{\partial x^\mu}(\frac{\partial L}{\partial z^i_\mu})= 0
\end{equation}
along $j^1 \sigma$. Equations (\ref{EL}) are called the {\sl Euler-Lagrange
equations} for $L$.

Next, we will recall the multisymplectic formulation
of classical field theories.

First of all, using the volume form $\eta$
we can construct a tensor field of type $(1,n)$ on $Z$: 
$$
S_{\eta} = (dy^i - z^i_{\mu} dx^{\mu}) \wedge d^{n-1}x^{\nu} \otimes
\frac{\partial}{\partial z^i_{\nu}}.
$$
The {\sl Poincar\'e-Cartan $n$ and $(n+1)$-forms} are defined as
$$
\Theta_L = \Lambda + S_{\eta}^* (dL), \qquad
\Omega_L = - d \Theta_L,
$$
where $S_{\eta}^*$ is the adjoint operator. 

\begin{proposition}{Proposition}
The $(n+1)$-form $\Omega_L$ is multisymplectic if and only if
the lagrangian $L$ is regular.
\end{proposition}

The integral action reads now as
$$
A(\sigma) = \int_U \, (j^1\sigma)^* \Theta_L.
$$

The use of the Poincar\'e-Cartan $(n+1)$-form
allows us to characterize the extremals as follows.

\begin{proposition}{Proposition}\cite{BSF}
$\sigma$ is an extremal for the action if and only if
\begin{equation}\label{fieldeq}
(j^1\sigma)^* (i_{\xi_Z} \, \Omega_L) = 0
\end{equation}
for all vector fields $\xi_Z$ on $Z$.
\end{proposition}
Eq. (\ref{fieldeq}) gives the {\sl field equations}, but we can consider a more
general problem: to look for a section $\tau$ of the fibration
$\pi_{XZ} : Z \longrightarrow X$ such that
\begin{equation}\label{dedondereq}
\tau^* (i_{\xi_Z} \, \Omega_L) = 0
\end{equation}
for all vector fields $\xi_Z$ on $Z$.
Eq. (\ref{dedondereq}) is referred as the {\sl De Donder equations}.

Looking the De Donder problem, one realizes that we search for sections
$\tau$ of $\pi_{XZ}$ satisfying Eq. (\ref{dedondereq}). We will linearize
the problem as follows.
First, notice that
the subspace $H_{\tau(x)}=d\tau(x)(T_xX) \subset T_{\tau(x)}Z$
is horizontal, for every $x$ in the domain of $\tau$. 
Indeed, we have
$$
T_{\tau(x)}Z = H_{\tau(x)} \oplus (V\pi_{XZ})_{\tau(x)},
$$
where $V\pi$ denotes the vertical bundle. 
Therefore, these horizontal subspaces can be considered
as infinitesimal approximations to the section in the same
way that tangent vectors are the linear approximations to the 
trajectories in particle mechanics.

Indeed, consider an arbitrary Ehresmann connection $\Gamma$ in the
fibration $\pi_{XZ} : Z \longrightarrow X$, that is, we have an
horizontal distribution $H$ which is complementary of the vertical
one:
$$
TZ = H \oplus V\pi_{XZ}.
$$
Denote by ${\bf h}$ the horizontal projector ${\bf h} : TZ \longrightarrow H$.
We then have
$$
{\bf h}(\frac{\partial}{\partial x^\mu}) = \frac{\partial}{\partial x^\mu}
+ \Gamma^i_{\mu} \frac{\partial}{\partial y^i} +
\Gamma^i_{\nu \mu} \frac{\partial}{\partial z^i_\nu}.
$$

A direct computation shows that
\begin{equation}\label{geomeq}
i_{\bf h} \, \Omega_L = (n-1) \, \Omega_L
\end{equation}
if and only if
\begin{eqnarray}\label{uno}
(\Gamma^j_{\nu} - z^j_{\nu}) 
\left(\frac{\partial^2 L}{\partial z^i_\mu \partial z^j_\nu}\right) = 0 \\
\label{dos}
\frac{\partial L}{\partial y^i} - \frac{\partial^2 L}{\partial x^\mu \partial z^i_\mu}
- \Gamma^j_\mu \frac{\partial^2L}{\partial y^j \partial z^i_\mu}
- \Gamma^j_{\mu \nu} \frac{\partial^2 L}{\partial z^j_\mu \partial z^i_\nu} +
(\Gamma^j_{\nu} - z^j_{\nu}) \frac{\partial^2 L}{\partial y^i \partial z^j_\nu} = 0
\end{eqnarray}
(see \cite{LMM}).

If the lagrangian $L$ is regular, then Eq. (\ref{uno}) implies that
$\Gamma^i_\mu = z^i_\mu$ and therefore (\ref{dos}) becomes
\begin{equation}\label{tres}
\frac{\partial L}{\partial y^i} - \frac{\partial^2 L}{\partial x^\mu \partial z^i_\mu}
- z^j_\mu \frac{\partial^2L}{\partial y^j \partial z^i_\mu}
- \Gamma^j_{\mu \nu} \frac{\partial^2 L}{\partial z^j_\mu \partial z^i_\nu}
= 0.
\end{equation}
Now, if $\tau(x^\mu)=(x^\mu, \tau^i(x), \tau^i_\mu(x))$ is an
integral section of $H$ we would have
$$
z^i_\mu=\frac{\partial \tau^i}{\partial x^\mu} \; ,
\Gamma^i_{\mu \nu}=\frac{\partial \tau^i_\mu}{\partial x^\nu}
$$
which proves that Eq. (\ref{tres}) is nothing but the Euler-Lagrange equations
for $L$.

As we have noticed, for regular lagrangians the field and De Donder
equations are equivalent, that is, a solution $\tau$ of the De Donder problem
is automatically a jet prolongation of a solution of
the field equations.

Eq. (\ref{geomeq}) can be considered as the linear version of De Donder equations.
If the lagrangian $L$ is regular, a global solution always exists, because one
can choose local solutions and use then a partition of the unity to glue together
all the local solutions. But this solution will not be unique.
Indeed, given a particular solution $\Gamma$ then any other solution
is of the form ${\bf h}+T$, where 
$\displaystyle{T=T^i_{\mu \nu} \, dx^\nu \otimes \frac{\partial}{\partial z^i_\mu}}$ 
is a tensor field on $Z$ of type $(1,n)$
satisfying the following conditions
\[
T^i_{\mu \nu} \frac{\partial^2 L}{\partial z^i_\mu \partial z^j_\nu} = 0, \;
\hbox{for all} \; j.
\]

\subsection{Presymplectic formalism in the space of Cauchy data}

Next, we will show how the selection of a Cauchy surface permits
to give an infinite dimensional setting for field theories
(see \cite{BSF} for a detailed exposition).

A {\sl Cauchy surface} is a compact oriented $(n-1)$-dimensional submanifold $B$
of $X$. We shall consider a manifold of embeddings from $B$ into $X$, denoted
by $\tilde{X}$; it will be the {\sl space of parametrized Cauchy surfaces}.

A space of {\sl Cauchy data} will be a manifold $\tilde{Z}$
of embeddings from $B$ into $Z$ such that $\pi_{XZ} \circ \gamma \in \tilde{X}$,
for all embedding $\gamma \in \tilde{Z}$.
This condition permits to define a fibration $\pi_{\tilde{X} \tilde{Z}} : \tilde{Z}
\longrightarrow \tilde{X}$.

A space of {\sl Dirichlet data} is a manifold $\tilde{Y}$ of embeddings from $B$ to $Y$
such that $\pi_{YZ} \circ \gamma \in \tilde{Y}$, for all $\gamma \in \tilde{Z}$.
Therefore we have a fibration $\pi_{\tilde{Y} \tilde{Z}} : \tilde{Z}
\longrightarrow \tilde{Y}$ such that
$\pi_{\tilde{X} \tilde{Z}} = \pi_{\tilde{X}\tilde{Y}} \circ \pi_{\tilde{Y}\tilde{Z}}$.

We define a 1-form $\tilde{\Theta}$ on $\tilde{Z}$ as follows:
\[
\tilde{\Theta}(\gamma)(\xi) = \int_B \, \gamma^*(i_{\xi} \Theta_L)
\]
where $\gamma \in \tilde{Z}$, and $\xi \in T_{\gamma}\tilde{Z}$. Notice that such a
tangent vector $\xi$ can be viewed as a mapping $\xi : B \longrightarrow TZ$
covering $\gamma$.

\begin{proposition}{Proposition}\cite{BSF}
We have
\[
d \tilde{\Theta} (\gamma)(\xi_1, \xi_2) = -
\int_B \, \gamma^* (i_{\xi_2} i_{\xi_1} \, \Omega_L)
\]
for all tangent vectors $\xi_1, \xi_2 \in T_{\gamma} \tilde{Z}$ and
for all $\gamma \in \tilde{Z}$.
\end{proposition}

Let $c_{\tilde{X}}(t)$ be a curve in $\tilde{X}$. We say that 
$c_{\tilde{X}}(t)$ {\sl splits} $X$ if the mapping
\[
(b, t) \mapsto c_{\tilde{X}}(t)(b)
\]
is a diffeomorphism from $B \times \R$ into $X$. (For simplicity, we are
assuming that the curves are defined in the whole real line).

Let now $c_{\tilde{Z}}(t)$ be a curve in $\tilde{Z}$ such that its projection
$\pi_{\tilde{X} \tilde{Z}} \circ c_{\tilde{Z}}(t)$ splits $X$. Under this condition, one can
define a section $\tau : X \longrightarrow Z$ of $\pi_{XZ}$ as follows:
\[
\tau(x) = c_{\tilde{Z}}(t)(b)
\]
since $x = \pi_{XZ} \circ c_{\tilde{Z}}(t)(b)$.

Conversely, given a section $\tau$ of $\pi_{XY}$ and a curve
$c_{\tilde{X}}(t)$ in $\tilde{X}$, we define a curve
$c_{\tilde{Z}}(t)$ in $\tilde{Z}$ by the formula
\[
c_{\tilde{Z}}(t)(b) = \tau(c_{\tilde{X}}(t)(b)).
\]

\begin{theorem}{Theorem}\cite{BSF}
Let $\tau$ be a section of $\pi_{XZ}$, and $c_{\tilde{X}}(t)$ a curve in $\tilde{X}$ 
such that $c_{\tilde{X}}(t)(B)$ is contained in the domain of $\tau$.
Let $c_{\tilde{Z}}(t)$ be the lift of $c_{\tilde{X}}(t)$ to $\tilde{Z}$ as we described above.
If $\tau$ satisfies the De Donder equation then
\begin{equation}\label{dynamics}
i_{\dot{c}_{\tilde{Z}}(t)} \, d\tilde{\Theta} = 0.
\end{equation}
Conversely, if the curve $c_{\tilde{Z}}(t)$ satisfies Eq. (\ref{dynamics})
and its projection splits $X$, then the section $\tau$ constructed as above
satisfies the De Donder equations.
\end{theorem}

\section{Classical field theories with non-holonomic constraints}

A non-holonomic classical field theory consists of a
lagrangian function $L=L(x^\mu, y^i, z^i_\mu)$ subjected
to constraints of the form
$$
\Phi^\alpha (x^\mu, y^i, z^i_\mu) = 0, \, 1 \leq \alpha \leq k.
$$
This means that the solutions of the field equations have to
satisfy the above equations along the full evolution of the system.

\begin{definition}{Definition}
A section $\sigma$ of $\pi_{XY} : Y \longrightarrow X$ is a solution
of the nonholonomic problem if (\ref{cero}) holds
for all vector fields $\xi_Y$ on $Y$ satisfying the condition
\begin{equation}\label{condition}
\frac{\partial \Phi^\alpha}{\partial z^i_\mu} \, \xi^i = 0, \qquad \alpha = 1, \dots, k.
\end{equation}
\end{definition}

\begin{remark}{Remark}
{\rm The above definition could be considered as a sort of d'Alembert principle
for constrained field theories. In principle, we consider
non-linear constraints. At the end of this section we will
treat the particular case of linear or affine constraints.}
\end{remark}

Proceeding as in the case of unconstrained systems we deduce
that $\sigma$ is a solution of the nonholonomic problem if and only if
$$
\int_U \, \left[\frac{\partial L}{\partial y^i} - 
\frac{\partial}{\partial x^\mu}(\frac{\partial L}{\partial z^i_\mu})\right]
\, \xi^i \, dx^n = 0
$$
for all values of $\xi^i$ satisfying (\ref{condition}) (here $U$ is
the domain of definition of $\sigma$). 

Therefore, such a solution $\sigma(x^\mu) = (x^\mu, \sigma^i(x))$
would satisfy the following set of equations:
\begin{eqnarray}\label{nonholoEL}
\frac{\partial L}{\partial y^i} - 
\frac{\partial}{\partial x^\mu}(\frac{\partial L}{\partial z^i_\mu}) & = &
\lambda_{\alpha \mu} \, \frac{\partial \Phi^\alpha}{\partial z^i_\mu}\\
\Phi^\alpha (x^\mu, \sigma^i, \frac{\partial \sigma^i}{\partial x^\mu}) & = & 0
\end{eqnarray}
along $j^1 \sigma$. Equations (\ref{nonholoEL}) are called the {\sl non-holonomic
field equations} for $L$. Here, the $\lambda_{\alpha \mu}$
are some Lagrange multipliers to be obtained.

\bigskip

We will rewrite the non-holonomic problem in a geometrical manner.

First of all, the constraints geometrically define a submanifold $M$ of $Z$.
Therefore, a field theory with non-holonomic constraints is described by
a lagrangian form $\Lambda = L \eta$ and a submanifold $M$ of $Z$.

The {\sl non-holonomic De Donder equations} can be then written as follows:
\begin{equation}\label{nonhologeom}
i_{\bf h} \, \Omega_L - (n-1) \Omega_L \in {\cal Y}(S_{\eta}^*(TM^o)), \qquad
\hbox{Im} \; {\bf h} \subset TM
\end{equation}
where ${\bf h}$ is the horizontal projector of a Ehresmann connection
$\Gamma$ in $\pi_{XZ} : Z \longrightarrow X$, $TM^o$ is the
annihilator of $TM$, and ${\cal Y}(S_{\eta}^*(TM^o))$
denotes the ideal generated by the space of $n$-forms
$S_{\eta}^*(TM^o)$.

We will check that, in fact, the above equations are correct.
Indeed, the first equation in (\ref{nonhologeom}) can be written as
\begin{equation}\label{otra}
i_{\bf h} \, \Omega_L - (n-1) \Omega_L = \lambda_\alpha \wedge 
S_{\eta}^*(d\Phi^\alpha), 
\end{equation}
with some 1-forms $\lambda_\alpha$. 

In the following, we will assume the following condition:

\medskip

{\bf Assumption on the regularity of the constraints}: 

{\sl The $n$-forms
$$
\{ S_{\eta}^*(d\Phi^\alpha) \; | \; \alpha = 1, \dots, k\}
$$
are linearly independent.}

\medskip 

Under this regularity assumption, it is esay to see that
$$
\lambda_\alpha = \lambda_{\alpha \mu} \, dx^\mu.
$$
Thus, if we introduce the notation
$$
\theta_\mu = \lambda_{\alpha \mu} \, S_{\eta}^*(d \Phi^\alpha)
$$
we deduce that (\ref{otra}) can be written as follows:
\begin{equation}\label{otra2}
i_{\bf h} \, \Omega_L - (n-1) \Omega_L = dx^\mu \wedge \theta_\mu .
\end{equation}

Now, from (\ref{nonhologeom}) we obtain
\begin{eqnarray}\label{unoligado}
(\Gamma^j_{\nu} - z^j_{\nu}) 
\left(\frac{\partial^2 L}{\partial z^i_\mu \partial z^j_\nu}\right) & = & 0\\
\label{dosligado}
\frac{\partial L}{\partial y^i} - \frac{\partial^2 L}{\partial x^\mu \partial z^i_\mu}
- \Gamma^j_\mu \frac{\partial^2L}{\partial y^j \partial z^i_\mu}
- \Gamma^j_{\mu \nu} \frac{\partial^2 L}{\partial z^j_\mu \partial z^i_\nu} +
(\Gamma^j_{\nu} - z^j_{\nu}) \frac{\partial^2 L}{\partial y^i \partial z^j_\nu} & = & 
\lambda_{\alpha \mu} \, \frac{\partial \Phi^\alpha}{\partial z^i_\mu}
\end{eqnarray}

If the lagrangian $L$ is regular, we deduce from 
(\ref{unoligado}) that $\Gamma^j_\mu = z^j_\mu$, and consequently, (\ref{dosligado})
becomes
\begin{equation}\label{nueva}
\frac{\partial L}{\partial y^i} - 
\frac{\partial}{\partial x^\mu}(\frac{\partial L}{\partial z^i_\mu}) = 
\lambda_{\alpha \mu} \, \frac{\partial \Phi^\alpha}{\partial z^i_\mu}.
\end{equation}
If $\tau(x^\mu)=(x^\mu, \tau^i(x), \tau^i_\mu(x))$ is an
integral section of $H$ we would have
$$
z^i_\mu=\frac{\partial \tau^i}{\partial x^\mu} \; , \qquad
\Gamma^i_{\mu \nu}=\frac{\partial \tau^i_\mu}{\partial x^\nu}
$$
and thus Eq. (\ref{nueva}) gives the non-holonomic field equations for $L$.

\medskip

Next we will give the infinite dimensional setting for non-holonomic
field theory.

To do that, we prove the following result.

\begin{proposition}{Proposition}
Let ${\bf h}$ be the horizontal projector of an Ehresmann connection
in the fibration $\pi_{XZ} : Z \longrightarrow X$, and let 
$\tau : X \longrightarrow Z$ be an integral section of this connection.
If ${\bf h}$ is a solution of the non-holonomic De Donder equations,
then we have
\begin{equation}\label{nuevados}
\tau^*(i_{\xi_Z} \, (\Omega_L - dx^\mu \wedge \theta_\mu)) = 0,
\end{equation}
for all vector fields $\xi_Z$ on $Z$. Conversely, if for all
the integral sections $\tau$ of ${\bf h}$ Eq. (\ref{nuevados}) holds
then ${\bf h}$ is a solution of the non-holonomic De Donder equations.
\end{proposition}

Eq. (\ref{nuevados}) can be viewed as the non-holonomic version
of the De Donder problem.

Choose now a Cauchy surface $B$ and the corresponding spaces
$\tilde{X}$, $\tilde{Y}$ and $\tilde{Z}$. It should
be noticed that now the space
of Cauchy data $\tilde{Z}$ consists of embeddings $\gamma : B \longrightarrow Z$
taking values in $M$.

As above we integrate the Poincar\'e-Cartan $(n+1)$-form
$\Omega_L$ to construct a 2-form $d\tilde{\Theta}$.
We will proceed in the same way with the constraints.
Let $\Xi$ be the $(n+1)$-form on $Z$ defined by
$$
\Xi = dx^\mu \wedge \theta_\mu.
$$
We then define a 2-form $\tilde{\Xi}$ on $\tilde{Z}$ as follows:
$$
\tilde{\Xi}(\gamma)(\xi_1, \xi_2) = -
\int_B \, \gamma^*(i_{\xi_2} i_{\xi_1} \, \Xi),
$$
for all $\gamma \in \tilde{Z}$ and for all tangent vectors
$\xi_1, \xi_2 \in T_{\gamma}\tilde{Z}$.

\begin{theorem}{Theorem}
Let $\tau$ be a section of $\pi_{XZ}$ taking values in $M$,
and $c_{\tilde{X}}(t)$ a curve in $\tilde{X}$ 
such that $c_{\tilde{X}}(t)(B)$ is contained in the domain of $\tau$.
Let $c_{\tilde{Z}}(t)$ be the lift of $c_{\tilde{X}}(t)$ to $\tilde{Z}$ as we described 
in the preceding section.
If $\tau$ satisfies the non-holonomic De Donder equation (\ref{nuevados}) then
\begin{equation}\label{nhdynamics}
i_{\dot{c}_{\tilde{Z}}(t)} \, (d\tilde{\Theta} - \tilde{\Xi}) = 0.
\end{equation}
Conversely, if the curve $c_{\tilde{Z}}(t)$ satisfies Eq. (\ref{nhdynamics})
and its projection splits $X$, then the section $\tau$ constructed as above
satisfies the non-holonomic De Donder equations.
\end{theorem}

{\bf Proof:} The proof follows the same lines that in \cite{BSF} for the unconstrained
case. Indeed, if $\tau$ satisfies (\ref{nuevados}), then we have
\begin{eqnarray*}
(i_{\dot{c}_{\tilde{Z}}(t)} \, d \Theta)(\xi_2) & = & - \int_B \, c_{\tilde{Z}}(t)^{*}
(i_{\xi_2} i_{\dot{c}_{\tilde{Z}}(t)}  \Omega_L)\\
\medskip
& = &  - \int_B \, c_{\tilde{Z}}(t)^*
(i_{\xi_2} i_{\dot{c}_{\tilde{Z}}(t)} \Xi)\\
\medskip
& = &  \tilde{\Xi}(c_{\tilde{Z}}(t))
(\dot{c}_{\tilde{Z}}(t), \xi_2)\\
\medskip
& = &  (i_{\dot{c}_{\tilde{Z}}(t)} \tilde{\Xi})(\xi_2)
\end{eqnarray*}
which implies (\ref{nhdynamics}).

The converse is proved by reversing the argument. \rule{5pt}{5pt}

\section{Affine and linear constraints}

Let ${\cal E}$ be a subspace of $\Lambda^n_2Y$. 
As we know, an element $z = j^1_x \sigma \in Z$ can be viewed
as an injective linear mapping $z=d\sigma(x) : T_xX \longrightarrow 
T_{\sigma(x)}Y$. Therefore, we could define a submanifold $M$ of $Z$
as follows:
$$
M = \{ z \in Z \; | \; d\sigma(x)(T_xX) \; \hbox{is annihilated by}
\; {\cal E} \}.
$$
We will prove that $M$ is the zero set of a family of affine constraints.

Indeed, take a basis 
$\{\varphi^\alpha \; , \alpha = 1, \dots, k\}$
of ${\cal E}$, and notice that each element
$\varphi^\alpha$ is locally expressed as
$$
\varphi^\alpha = (\varphi^\alpha)_0 \, d^nx +
(\varphi^\alpha)^\mu_i \, dy^i \wedge d^{n-1}x^\mu.
$$
Since
$$
d\sigma(x)(\frac{\partial}{\partial x^\mu}) =
\frac{\partial}{\partial x^\mu} + z^i_\mu \,
\frac{\partial}{\partial y^i},
$$
where $z=(x^\mu, y^i, z^i_\mu)$, we have
$$
\varphi^\alpha (d\sigma(x) \frac{\partial}{\partial x^1}, \dots,
d\sigma(x) \frac{\partial}{\partial x^n}) =
(\varphi^\alpha)_0 + (\varphi^\alpha)^\mu_i z^i_\mu,
$$
which is an affine constraint. Linear constraints are obtained 
considering a subspace of the quotient vector bundle
$\displaystyle{
Z^* = \frac{\Lambda^n_2Y}{\Lambda^n_1Y}}$.

\section{An example: a pneumatic tire}

We shall briefly describe a mathematical model for
an automobile tire rolling with constant velocity, while deforming.

There are several mathematical models: Rocard's theory, Greidanu's theory and
Keldys's theory (see \cite{NF}, Ch. VI, pp. 308 and ff.)
The last one is the most interesting.
We shall describe it.

For a pneumatic tire, we introduce the following parameters describing a configuration:
the coordinate $x$ of the point $K$ of intersection of the diameter of greatest slope
of the wheel and the plane $Oxy$ of the road, the angle $\kappa$ between the normal
to the road and the central plane of the wheel,
the angle $\theta$ between the $Oy$ axis and the path of the central
plane of the wheel on the road, the lateral displacement $\xi$ of the center of the area
of contact of the tire with respect to $K$,
and the angle $\varphi$ measuring the torsional deformation of the tire
(see Figure).

\begin{wrapfigure}{l}{3.3cm}
\includegraphics[width=3.3cm,clip]{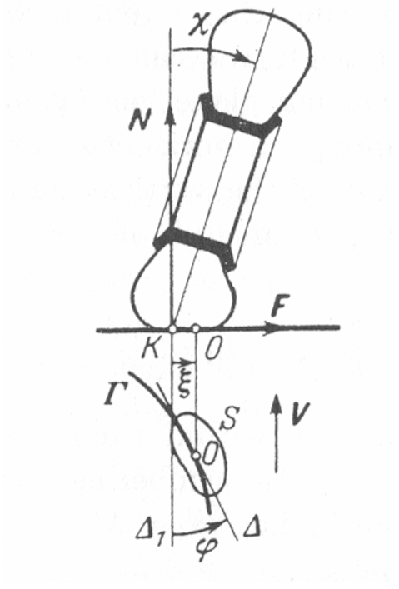}
\end{wrapfigure}

The lagrangian is
$$
L(x, \kappa, \theta, \xi, \varphi, \dot{x}, \dot{\kappa}, \dot{\theta}, \dot{\xi}, \dot{\varphi})=
T(x, \kappa, \theta, \dot{x}, \dot{\kappa}, \dot{\theta}) - U(\kappa, \xi, \varphi),
$$ 
where $T$ is the kinetic energy and $U$ is an elastic potential.
The kinetic energy of deformation of the tire is neglected, and the potential
energy is
$$
U(\xi, \phi, \kappa) = \frac{1}{2} \, 
\left( a \xi^2 + b \phi^2 + \rho N \kappa^2 + 2 \sigma N \xi \kappa \right),
$$
where $a, b, \sigma$, and $\rho$ are constants to be
experimentally determined, and $N$ is the normal load.
It should be noticed that when $N$ increases the area of
contact of the tire, which remains at rest during
the loading of the wheel, increases.

The kinematic constraints are
\begin{eqnarray*}
\dot{x} + \dot{\xi} + V \theta + V \phi &=& 0\\
\dot{\theta} + \dot{\varphi} - \alpha V \xi + \beta V \varphi
+ \gamma V \kappa &=& 0,
\end{eqnarray*}
where $\alpha, \beta$ and $\gamma$ are constants, and $V$ is the constant velocity
of the wheel.

The stability of vehicles on wheels with pneumatic tires is carefully analyzed
in \cite{NF}, introducing several simplifications to treat with the motion equations.

It should also be noticed that for accelerated motions,
the area of the contact of the tire with road is not more constant
(or even simply connected). This would lead to introduce unilateral
constraints. There are mathematical descriptions for particle mechanic
systems subjected to unilateral constraints (see \cite{madrid1,madrid2,brogliato} 
and the references therein)
but the extension to classical field theories is not developed yet.

\section{Concluding remarks and future work}

We have developed a multisymplectic setting for first-order
field theories subjected to non-holonomic constraints
which is very similar to that for particle mechanics (see \cite{LM}).
The dynamics in the corresponding space
of Cauchy data was also analyzed.

\subsection{Algebraic computation of the Lagrange multipliers}

In Particle Mechanics the Lagrange multipliers are obtained using
a simple algebraic method, indeed, the problem is reduced to apply
the Cramer rule for systems of linear equations \cite{LM}.
Therefore, one can use {\sc Mathematica}$^{\tiny \copyright}$ to do the computation.
A similar procedure can be used for field theories, but
the situation is really more complicated.
In addition, we are trying to obtain the constrained Ehresmann connection
by projecting a solution of the unconstrained problem.

\subsection{Applications to Continuum Mechanics}

As we said in the Introduction, the most interesting situations
occur in Continuun Mechanics.
The case of non-holonomic constraints given by connections
in media with microstructure are being analyzed in \cite{BLMS1}.
Some independent work can be found in
\cite{PL1,vignolo}.

\subsection{Non-holonomic bracket,
non-holonomic momentum mapping and reduction}

These topics will be analyzed in \cite{BLMS1} and \cite{LMS},
following the lines in \cite{CIL2} and \cite{CLMM}.

\section*{Acknowledgements}

The authors wish to thank the referee for his comments and remarks
that help to improve this paper.

\end{document}